\documentclass[runningheads,a4paper]{llncs}
\usepackage{amssymb}
\usepackage{amsmath}
\usepackage{graphicx}
\usepackage{float}
\usepackage{longtable}
\usepackage{cite}
\usepackage{caption}
\usepackage{subfig}

\title{Relation between Insertion Sequences and Genome Rearrangements in \textit{Pseudomonas aeruginosa}}
\titlerunning{Relation between Insertion Sequences and Bacterial Genome Rearrangements}
\author{
	Huda Al-Nayyef$^{1, 3}$, Christophe Guyeux$^1$, Marie Petitjean$^{2}$, Didier Hocquet$^{2}$ and Jacques M. Bahi$^1$\\
\authorrunning{Lecture Notes in Computer Science: Authors' Instructions}	
\institute{$^1$~FEMTO-ST Institute, UMR 6174 CNRS, DISC Computer Science Department \\
	Universit\'e de Franche-Comt\'e,
	16, route de Gray, 25000 Besan\c{c}on, France.
$^2$~Laboratoire d'Hygi\`ene Hospitali\`ere,  UMR 6249 CNRS Chrono-environnement, Universit\'e de Franche-Comt\'e, France.\\
$^3$~Computer Science Department, University of Mustansiriyah, Iraq.\\
\{huda.al-nayyef, christophe.guyeux, marie.petitjean, jacques.bahi\}@univ-fcomte.fr\\
dhocquet@chu-besancon.fr
}
}
\begin{document}

\mainmatter

\maketitle

\begin{abstract}
During evolution of microorganisms genomes underwork have different changes in their lengths, gene orders, and gene contents. Investigating these  structural rearrangements helps to understand how genomes have been modified over time. Some elements that play an important role in genome rearrangements are called insertion sequences (ISs), they are the simplest types of transposable elements (TEs) that widely spread within prokaryotic genomes. ISs can be defined as DNA segments that have the ability to move (cut and paste) themselves to another location within the same chromosome or not. Due to their ability to move around, they are often presented as responsible of some of these genomic recombination. Authors of this research work have regarded this claim, by checking if a relation between insertion sequences (ISs) and genome rearrangements can be found. To achieve this goal, a new pipeline that combines 
various tools
has firstly been designed, for detecting the distribution of ORFs that belongs to each IS category.
Secondly, links between these predicted ISs and observed rearrangements of two close genomes have been investigated, by seeing them with the naked eye,  and by using computational approaches.
The proposal has been tested on 18 complete bacterial genomes of \textit{Pseudomonas aeruginosa}, leading to the conclusion that IS3 family of insertion sequences are related to genomic inversions.

\keywords{Rearrangements, Inversions, Insertion Sequences, Pseudomonas aeruginosa}
\end{abstract}

\section{Introduction}
The study of genome rearrangements in microorganisms has become very important in computational biology and bio-informatics fields, owing to its applications in the evolution measurement of difference between species~\cite{lin2006spring}. 
Important elements in understanding genome rearrangements during evolution are called transposable elements, which are DNA fragments or segments that have the ability to insert themselves into new chromosomal locations, and often make duplicate copies of themselves during transposition process~\cite{nref1hawkins2006differential}. Indeed, within 
bacterial species, only cut-and-paste of transposition mechanism can be found, the 
transposable elements involved in such way being the insertion sequences. 
These types of mobile genetic elements (MGEs) seem to play an essential role in genomes rearrangements and evolution of prokaryotic genomes~\cite{Siguier2006526,citeulike:1766382}.

\begin{table}[H]
\center
\caption{P. aeruginosa isolates used in this study.}\label{tab:GenesNb}
\scalebox{0.7}{
\begin{tabular}{c|c|c}
Isolates & NCBI accession number & Number of genes  \\
\hline
19BR & 485462089 & 6218\\
213BR & 485462091 & 6184\\
B136-33 & 478476202 & 5818\\
c7447m & 543873856 & 5689 \\
DK2 & 392981410 & 5871\\
LES431 & 566561164 & 6006\\
LESB58 & 218888746 & 6059\\
M18 & 386056071 & 5771\\
MTB-1 & 564949884 & 6000\\
NCGM2.S1 & 386062973 & 6226\\
PA1 & 558672313 & 5981\\
PA7 & 150958624 & 6031\\
PACS2 & 106896550 & 5928\\
PAO1 & 110645304 & 5681\\
RP73 & 514407635 & 5804\\
SCV20265 & 568306739 & 6190\\
UCBPP-PA14 & 116048575 & 5908\\
YL84 & 576902775 & 5856\\ 
\end{tabular}
}
\end{table}

In this research work, we questioned the relation between the movement of insertion sequences on the one hand, and genome rearrangements on the other hand, and tested whether the type of IS family influences this relation. 
Investigations will focus on inversion operations of rearrangement (let us recall that an inversion occurs within genomes when a 
chromosome breaks at two points, and when the segment flanked with these breakpoints is inserted again but in reversed order, this event being potentially mediated with molecular mechanisms~\cite{kirkpatrick2010and,ranz2007principles}).
To achieve our goal, we built a pipeline system module that combines existing tools together with the development of new ones, for finding putative ISs and inversions within studied genomes. We will then use this system to investigate the structure of prokaryotic genomes, by searching for IS elements at the boundaries of each inversion. 

The contributions of this article can be summarized as follows. (1) 
A pipeline for insertion sequences discovery and classification is proposed. 
It uses unannotated genomes and then combines different existing tools for ORF predictions and clustering. It also classifies them according to an international IS database specific to bacteria. Involved tools in this stage are, among others, Prodigal~\cite{hyatt2010prodigal}, Markov Cluster Process (MCL)~\cite{van2000graph}, and ISFinder\footnote{\url{www-is.biotoul.fr}}~\cite{siguier2006isfinder}.
(2) We then use two different strategies to check the relation between ISs and genomic rearrangements. The first one used a well-supported phylogenetic tree, then genomes of close isolates are drawn together, while the questioned relation is checked with naked eye. In the second strategy, inversion cases are thoroughly investigated with ad hoc computer programs. And (3), the pipeline is tested on the set of 18 complete genomes of \textit{Pseudomonas aeruginosa} provided in Table~\ref{tab:GenesNb}. After having checked left and right inversion boundaries according to different window sizes, the probability of appearance of each type of IS family is then provided, and 
biological consequences are finally outlined.

The remainder of this article is organized as follows. 
The proposed pipeline for detecting insertion sequences in a list of ORFs extracted from unannotated genomes is detailed in Section~\ref{sec:methods}.
Rearrangements found using drawn genomes of close isolates is detailed in 
Section~\ref{sec:rearrange}, while a computational method for discovering inversions within all 18 completed genomes of \textit{P. aeruginosa} and results are provided in Section~\ref{sec:inversion}.
 This article ends by a conclusion section, in which the contributions are summarized and intended future work is detailed.

\begin{figure}[!hb]
\centering
\includegraphics[scale=0.5]{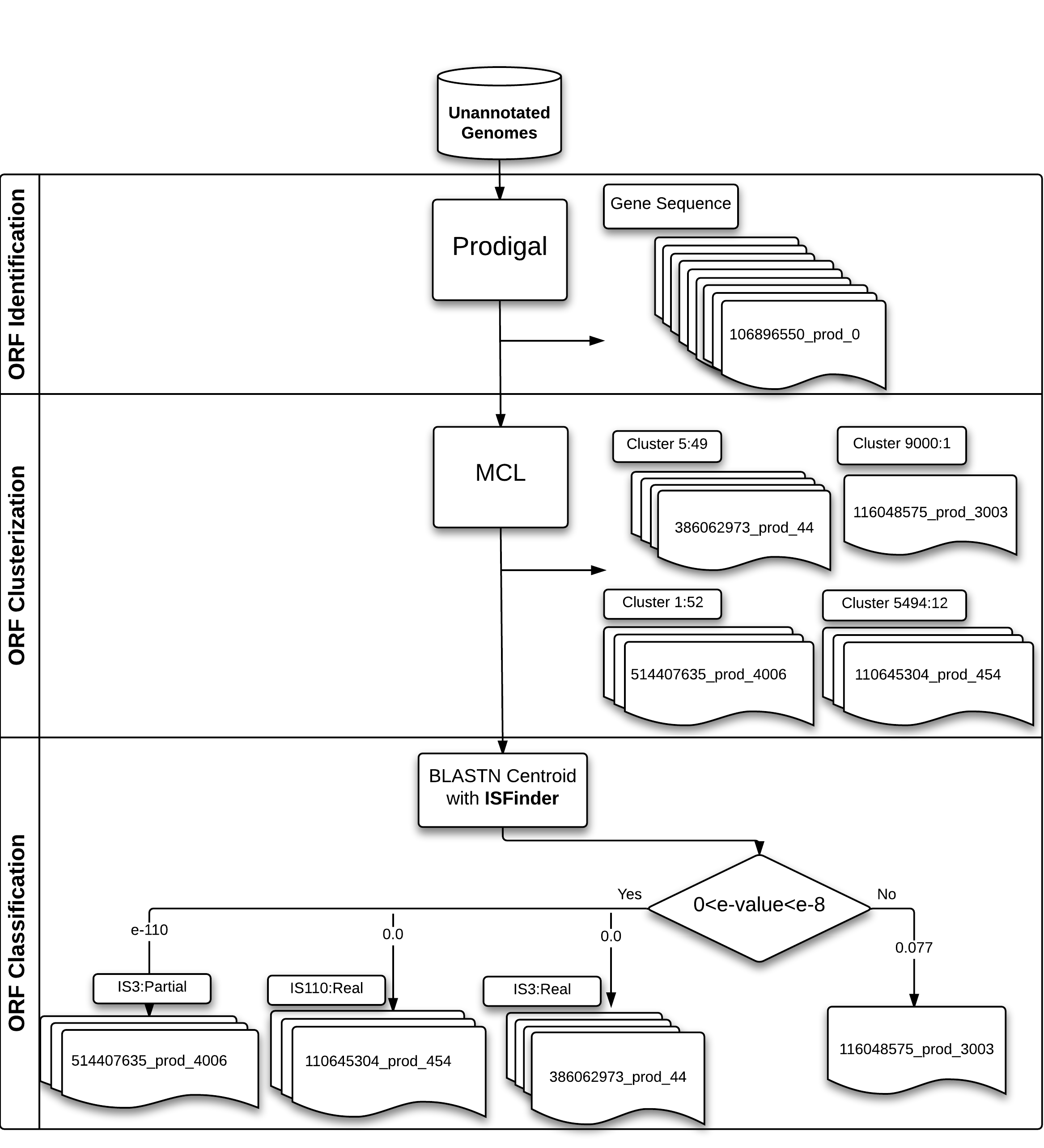}
\caption{Pipeline for detecting IS clusters in genomes of \textit{P.aeruginosa}}
\label{fig:IS_Module}
\end{figure}

\section{Methodology for IS detection}
\label{sec:methods}

In a previous work~\cite{huda2014}, we have constructed a pipeline system that combines three annotation tools 
(BASys~\cite{van2005basys}, Prokka~\cite{seemann2014prokka}, and 
Prodigal~\cite{hyatt2010prodigal}) with OASIS~\cite{robinson2012oasis}, 
that detected IS elements within prokaryotic 
genomes. This pipeline produces various information about each predicted IS, each IS is bordered by an \textit{Inverted-Repeat} (IR) sequence, number of ORFs in each IS family 
and group, etc. 
As we are now only interested in detecting which ORFs are insertion sequences, 
we have developed a new lightweight pipeline that focuses on such open reading frames..
This pipeline, depicted in  Figure~\ref{fig:IS_Module}, relies on ISFinder
 database~\cite{siguier2006isfinder}, the up-to-date reference for bacterial insertion sequences. The main function of this database is to assign IS names and to provide a focal point for a coherent nomenclature.
This is also a repository for ISs that contains all information about insertion sequences such as family and group. 

The proposed pipeline can be summarized as follows.
\begin{description}
\item[Step 1: ORF identification.] Prodigal is used as annotation tool for predicting gene sequences. This tool is an accurate bacterial and archaeal gene finding  software
provided by the Oak Ridge National Laboratory~\cite{hyatt2010prodigal}. Table~\ref{tab:GenesNb} lists the number of the predicted genes in each genome.
\item[Step 2: ORF clustering.] The Markov Cluster Process (MCL) algorithm is then used to achieve clustering of detected ORFs~\cite{van2000graph,enright2002efficient}.
\item[Step 3: Clusters classification.]

The IS family and group of the centroid sequences of each cluster is determined with ISFinder database.
BLASTN program is used here: if the e-value of the first hit is equal to 0, then the cluster of the associated sequence is called a ``Real IS cluster''. Otherwise, if the e-value is lower than $10^{-8}$, the cluster is denoted as ``Partial IS''.
At each time, family and group names of ISs that best match the considered sequence 
are assigned to the associated cluster. In Table~\ref{IS_info} summarizes founded IS clusters found in the 18 genomes of \textit{P. aeruginoza}. 
\end{description}

\begin{table}[H]
\centering
\caption{Summary of detected IS clusters}
\label{IS_info}
\begin{tabular}{l|c|c|c}
\hline
 & \multicolumn{1}{l|}{No. of Clusters} & \multicolumn{1}{l|}{Max. size of Cluster} & \multicolumn{1}{l}{Total no. of IS genes} \\ \hline
Partial IS & 94 & 57 & 362 \\ 
Real IS & 66 & 49 & 238 \\ \hline
Total IS Cluster & 160 & - & 600 \\ 
\hline
\end{tabular}
\end{table}

\section{Rearrangements in \textit{Pseudomonas aeruginosa}}
\label{sec:rearrange}
At the nucleotide level, genomes evolve with point mutations and small insertions and deletions~\cite{garcia2006mechanism},
 while at genes level, larger modifications including duplication, deletion, or inversion, of a single gene or of a large DNA segment, affect genomes by large scale rearrangements~\cite{hurles2004gene,proost2012adhore}. 
The pipeline detailed previously investigated the relations between insertion sequences and these genome rearrangements, by using two different methods that will be described below.

\subsection{Naked eye investigations}
\begin{figure}[H]
\centering
\includegraphics[scale=0.3]{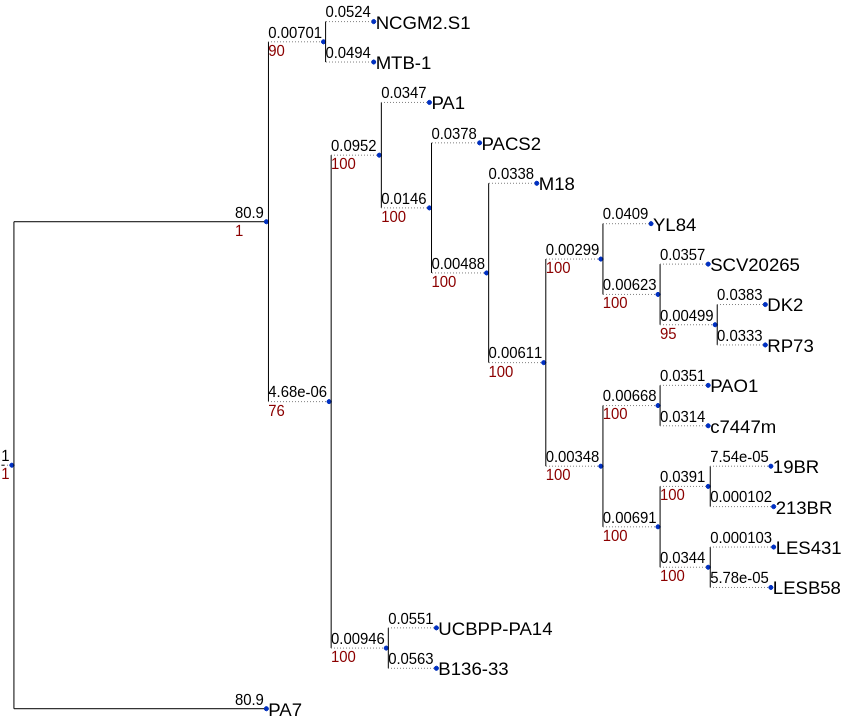}
\caption{Phylogeny of \textit{P. aeruginosa} based on mutations on core genome}
\label{fig:phylo}
\end{figure}

In order to visualize the positions of IS elements involved in genomic recombination that have occurred in the considered set of \textit{Pseudomonas}, 
we have first designed Python scripts that enable us to humanly visualize close genomes. Each complete genome has been
annotated using the pipeline described in the previous section, and the strict core genome has been extracted. This latter is constituted 
by genes shared exactly once in each genome. Thus polymorphic nucleotides in these core genes have been extracted, and a phylogeny using maximum likelihood (RAxML~\cite{Stamatakis21012014,alkindy2014finding} with automatically detected mutation model) has been inferred. (Figure~\ref{fig:phylo}).

For each close isolates, a picture has then been produced using our designed Python script, for naked eye investigations.
Real and Partial IS are represented with a red and green circles,respectively.
Additionally, DNA sequences representing the same gene have been linked either with a curve 
(two same genes in the same isolate) or with a line (two same genes in two close isolates).
Example of recombination events are given below.

\begin{figure}[H]
\begin{center}
    \subfloat[Insertion events of IS sequences have occurred in this set of 18 \textit{P. aeruginosa} species. For instance, when comparing DK2 and RP73, we have found that IS3-IS3 (2 ORFs) and IS3-IS407 (2 ORFs too) have been inserted inside RP73.\label{fig:ins}]{%
    \includegraphics[width=0.7\textwidth]{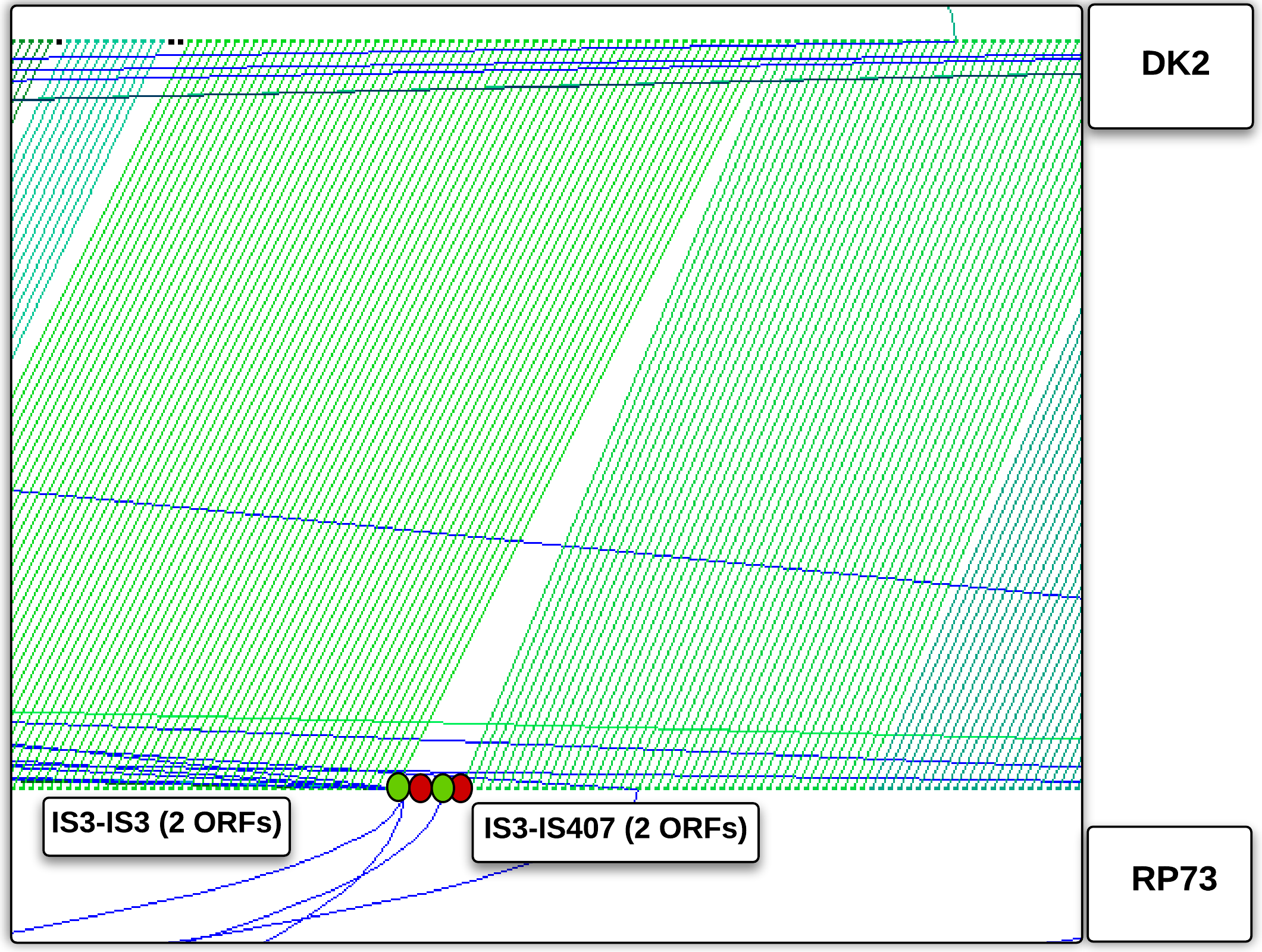} 
    }
    \vfill
    \subfloat[Deletions of insertion sequences can be found too, IS5 (Partial IS) is present in the genome of DK2, while it is deleted in the close isolates RP73.\label{fig:del}]{%
      \includegraphics[width=0.7\textwidth]{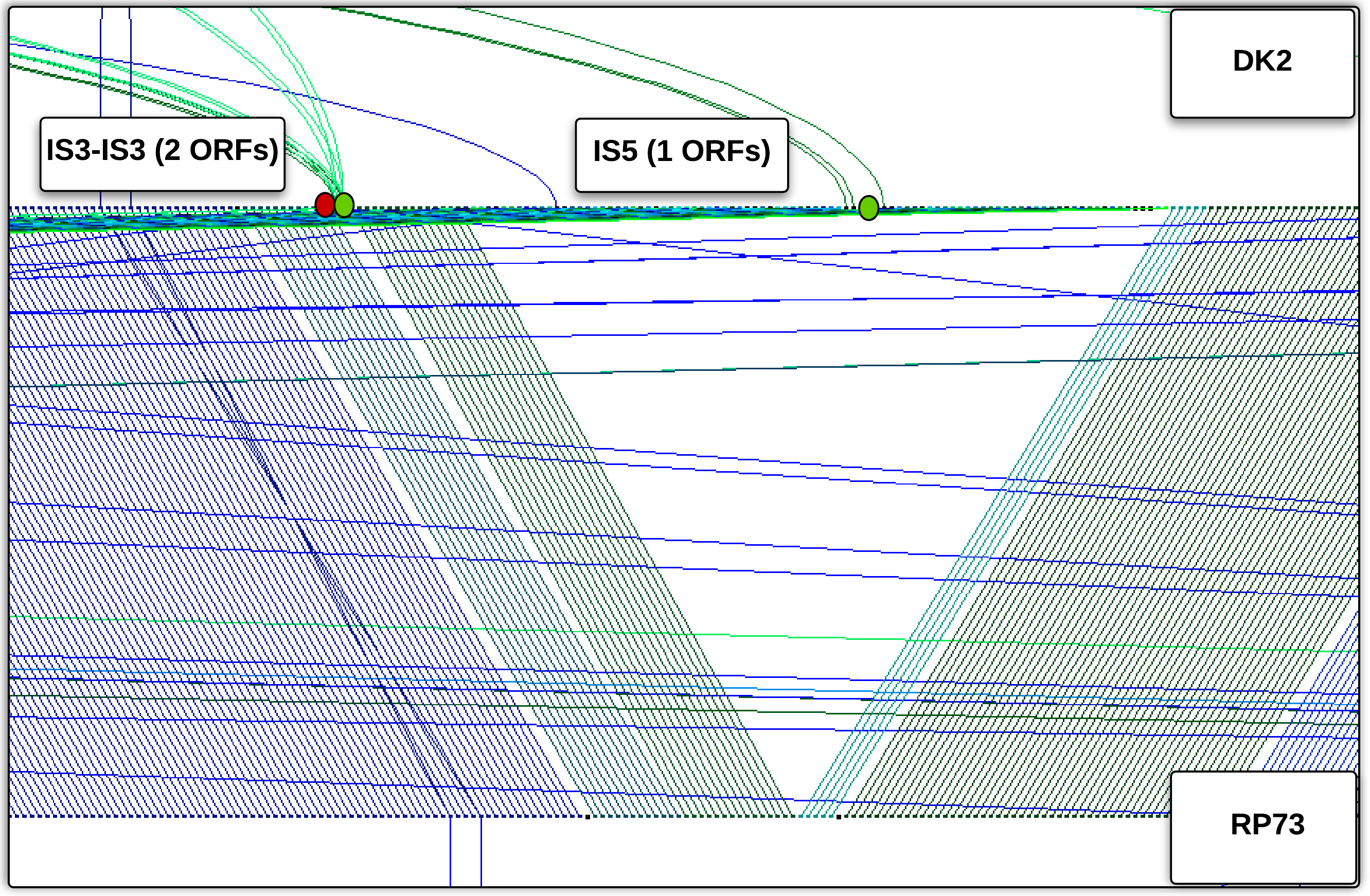} 
    }
    \vfill
    \subfloat[A duplication occurs in the insertion sequence type IS110-IS1111 that contains one ORF (Real IS), as there are 6 copies of this insertion sequence in both PAO1 and C7447m genomes.\label{fig:dup}]{%
      \includegraphics[width=0.7\textwidth]{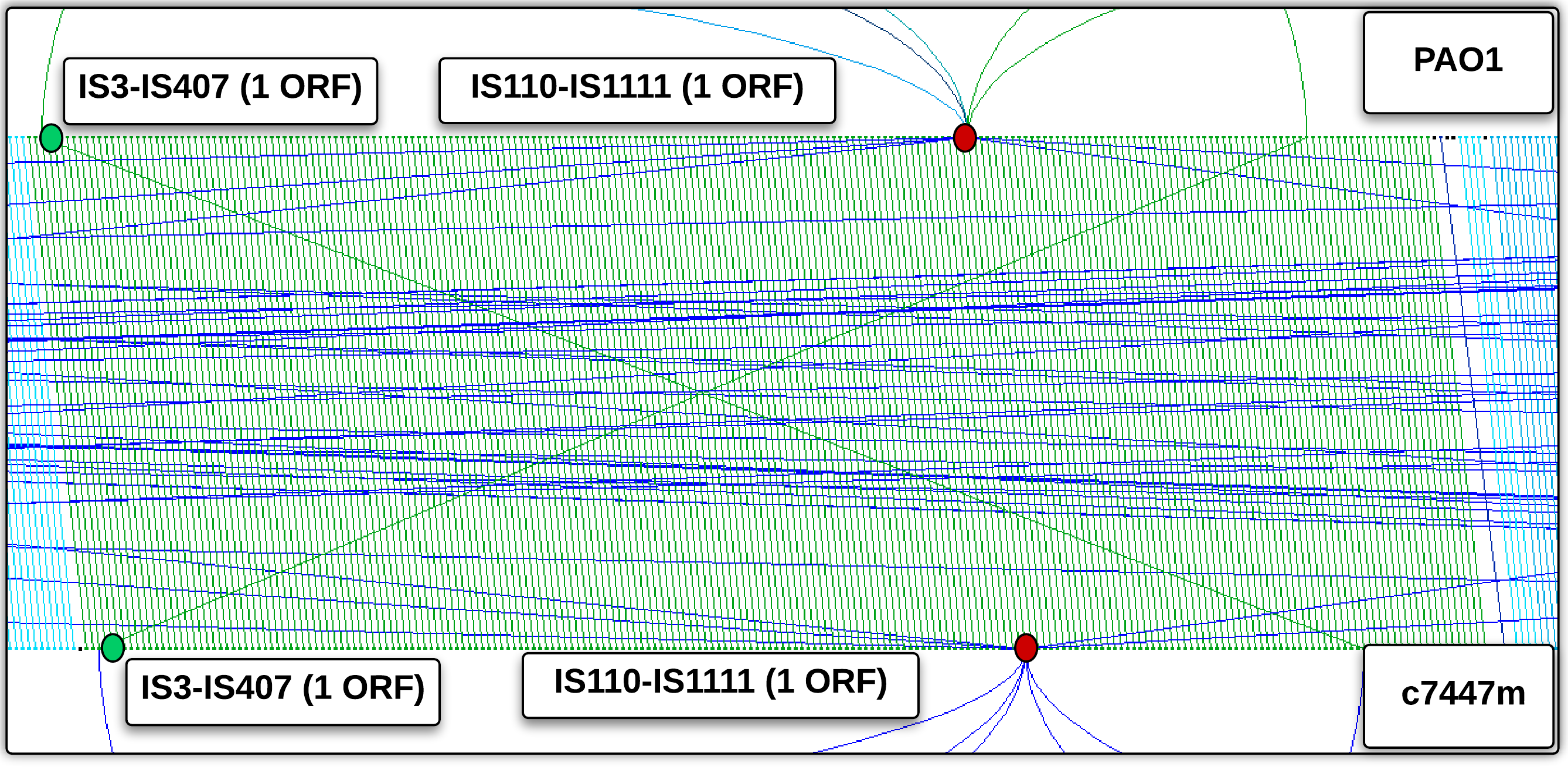} 
    }
    \caption{ Examples of genomic recombination events: Insertion, Deletion, and Duplication.}
    \label{fig:events}
\end{center}
\end{figure}
\begin{figure}[H]
\begin{center}
    \includegraphics[width=0.7\textwidth]{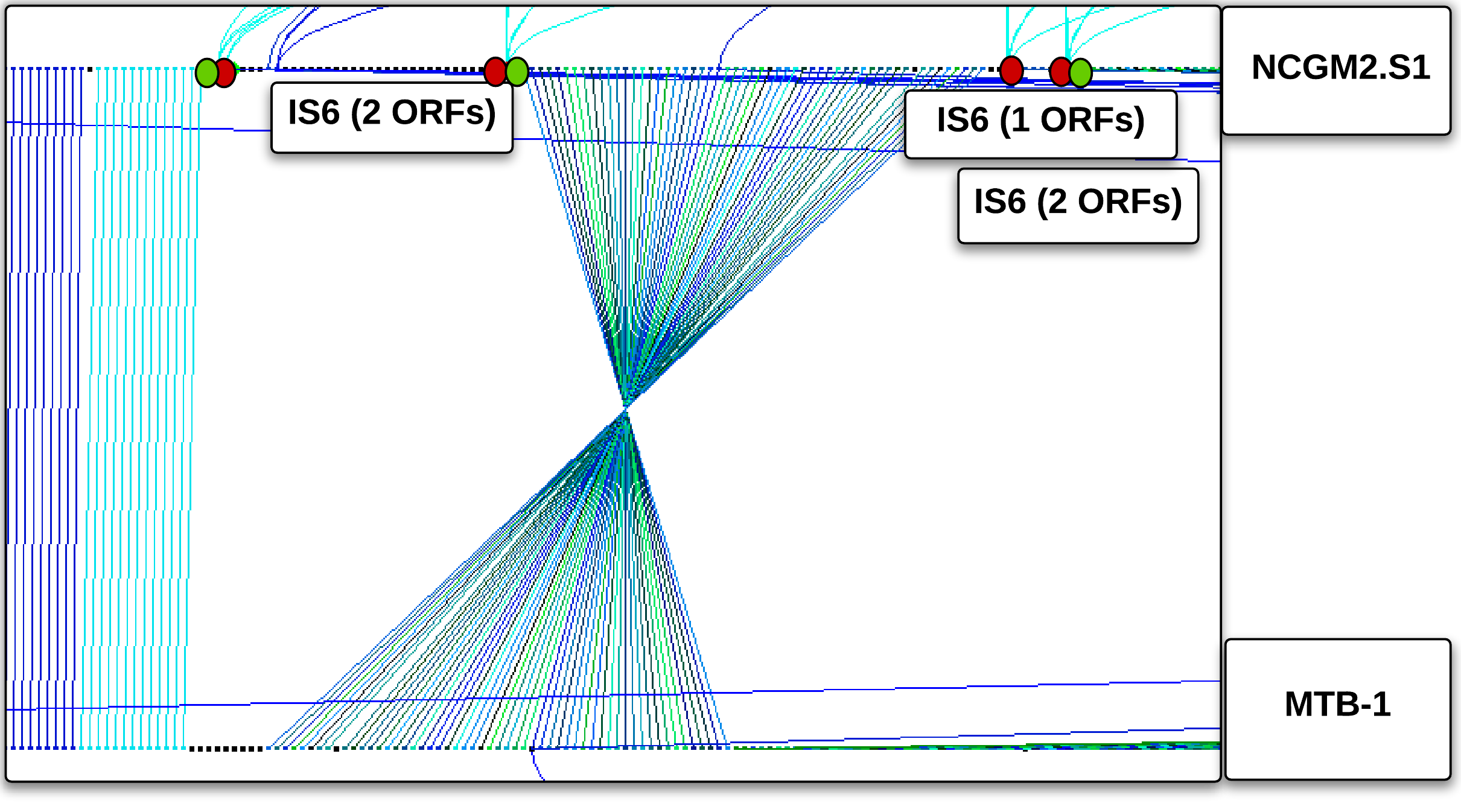} 
    \caption{The surrounding insertion sequences are within the same IS family (IS6) in the NCGM2.S1 genome. We have found too that insertion sequences are not always exactly at the beginning and end positions of the inversion, but they are overrepresented near these boundaries.}\label{fig:inv1}
\end{center}
\end{figure}

We will focus now on the link between large scale inversions and ISs as shown in Figure~\ref{fig:inv1}, by designing another pipeline that 
automatically investigate the inversions.

\subsection{Automated investigations of inversions}
\label{sec:inversion}
 
\begin{figure}[!ht]
\centering
\includegraphics[scale=0.52]{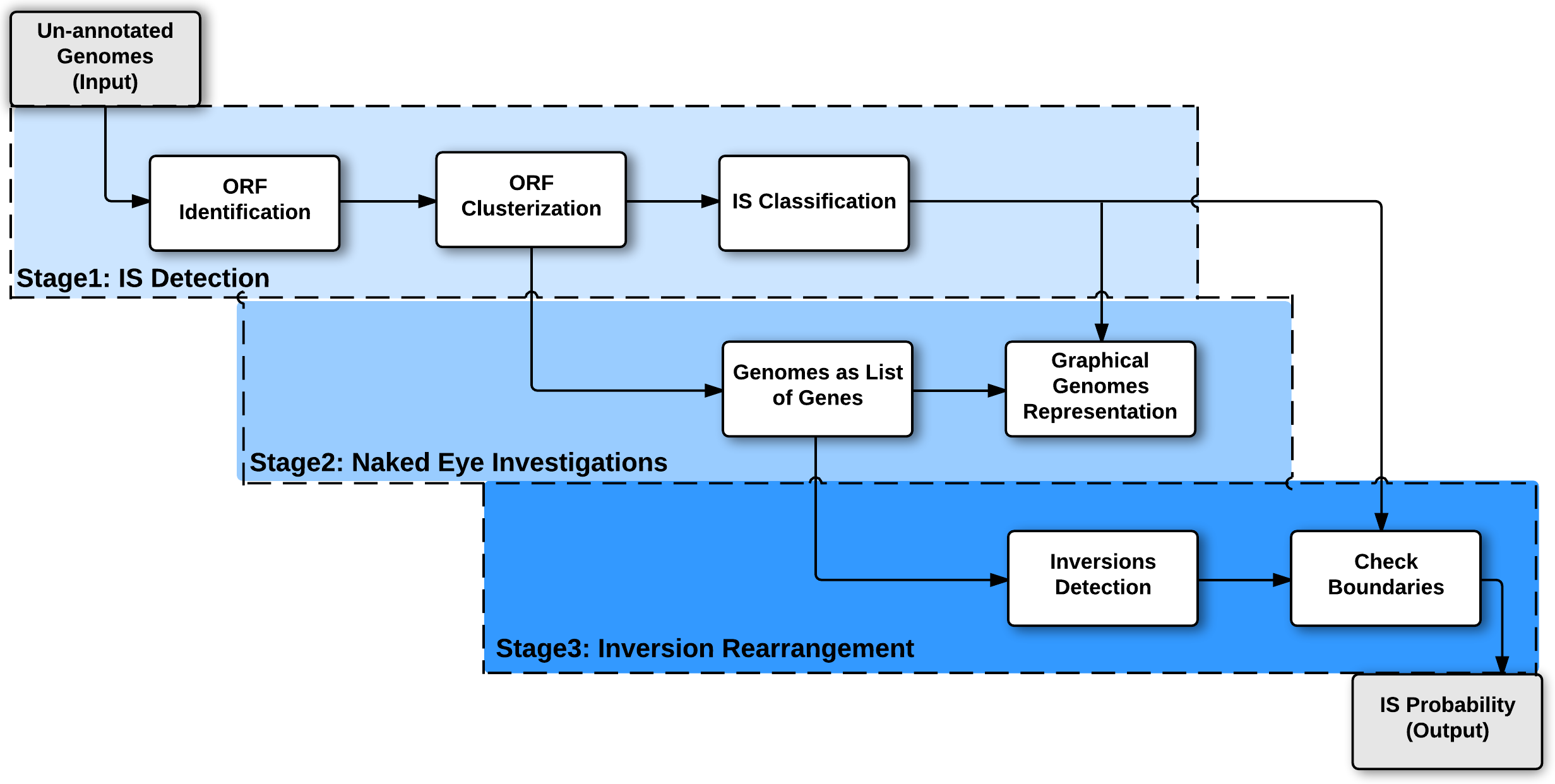}
\caption{Pipeline for detecting the role of ISs in inversions}
\label{fig:overall}
\end{figure}

The proposal is now to automatically extract all inversions that have occurred within the set of 18 genomes under consideration, and then to investigate their relation with predicted IS elements. The proposed pipeline is described in Figure~\ref{fig:overall}.
 
\begin{table}[!ht]
\center
\caption{Small and large inversions detected from all genomes}
\scalebox{0.7}{
\begin{tabular}{|c|c|c|c|c|c|c|}
\hline
\textbf{Genome 1} & \textbf{Start} & \textbf{Stop} & \textbf{Genome 2} & \textbf{Start} & \textbf{Stop} & \textbf{Length (genes)} \\ \hline
19BR & 1001 & 1002 & 213BR & 5094 & 5095 & 2 \\ 
19BR & 2907 & 2920 & 213BR & 2933 & 2946 & 14 \\ \hline
19BR & 684 & 685 & LES431 & 4689 & 4690 & 2 \\ 
19BR & 850 & 978 & LES431 & 4393 & 4521 & 129 \\ \hline
PAO1 & 997 & 998 & c7447m & 1977 & 1978 & 2 \\ \hline
LESB58 & 4586 & 4587 & LES431 & 2347 & 2348 & 2 \\ \hline
DK2 & 2602 & 2603 & RP73 & 2516 & 2517 & 2 \\ 
DK2 & 1309 & 1558 & RP73 & 3489 & 3738 & 250 \\ \hline
DK2 & 260 & 261 & SCV20265 & 3824 & 3825 & 2 \\ 
DK2 & 2846 & 2852 & SCV20265 & 3065 & 3071 & 7 \\ \hline
M18 & 3590 & 3591 & PACS2 & 1920 & 1921 & 2 \\ 
M18 & 3194 & 3579 & PACS2 & 2076 & 2461 & 386 \\ \hline
MTB-1 & 5581 & 5582 & B136-33 & 4742 & 4743 & 2 \\ \hline
UCBPP-PA14 & 4820 & 4821 & B136-33 & 2871 & 2872 & 2 \\ \hline
NCGM2.S1 & 1053 & 1307 & MTB-1 & 4507 & 4761 & 255 \\ 
NCGM2.S1 & 1742 & 1743 & MTB-1 & 4882 & 4883 & 2 \\ \hline
PA1 & 95 & 96 & B136-33 & 2691 & 2692 & 2 \\ 
PA1 & 1334 & 1491 & B136-33 & 1286 & 1443 & 158 \\ \hline
PACS2 & 94 & 97 & PA1 & 495 & 498 & 4 \\ 
PACS2 & 970 & 1206 & PA1 & 2220 & 2456 & 237 \\ \hline
SCV20265 & 45 & 46 & YL84 & 721 & 722 & 2 \\ 
SCV20265 & 261 & 462 & YL84 & 306 & 507 & 202 \\ \hline
UCBPP-PA14 & 259 & 260 & B136-33 & 3507 & 3508 & 2 \\ \hline
YL84 & 721 & 722 & M18 & 43 & 44 & 2 \\ 
YL84 & 768 & 983 & M18 & 5555 & 5770 & 216 \\ \hline
YL84 & 721 & 722 & PAO1 & 44 & 45 & 2 \\ 
YL84 & 1095 & 1264 & PAO1 & 5192 & 5361 & 170 \\ \hline
\end{tabular}
}
\label{long}
\end{table}

\begin{itemize}
\item \textbf{Step1}: Convert genomes from the list of predicted coding sequences in the list of integer numbers, by considering the cluster number of each gene. 
\item \textbf{Step2}: Extract sets of inversion from all input genomes. 719 inversions have been found (see Table~\ref{long}).

\item \textbf{Step 3}: Extract IS clusters (Partial and Real IS) using the first pipeline, as presented in a previous section.

\begin{figure}[!ht]
\centering
    \subfloat[Left and Right Boundary using window \label{fig:wind}]{%
    \includegraphics[width=0.5\textwidth]{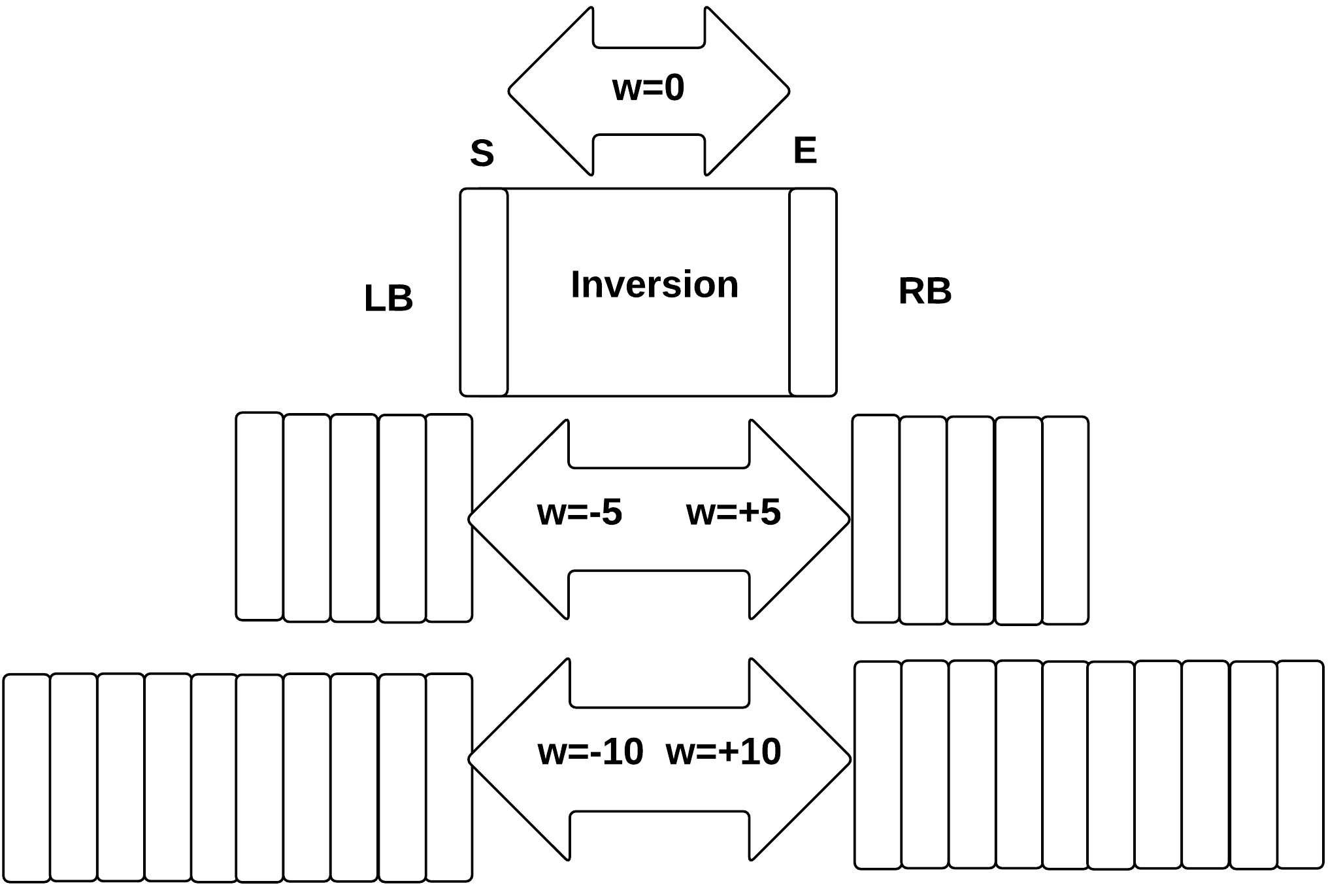}
    }
    \subfloat[No. of inversions for three different window size\label{fig:chartcases}]{%
      \includegraphics[width=0.6\textwidth]{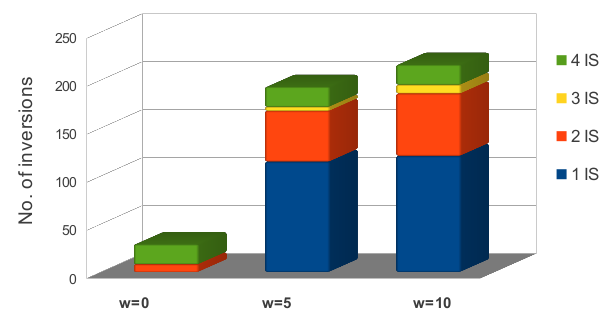}
    }
    
    \caption{Using different window size within all inversions}
    \label{fig:wind_cases}
\end{figure}

\item \textbf{Step 4}: Investigate boundaries of each inversion (starting S and ending E positions), by checking the presence of insertion sequences within a window ranging from $w=0$ up to 10 genes. Between 0 and 4 insertion sequences have been found at the boundaries of each inversion, (Figure~\ref{fig:wind_cases}).

\begin{figure}[!ht]
\centering
\includegraphics[scale=0.35]{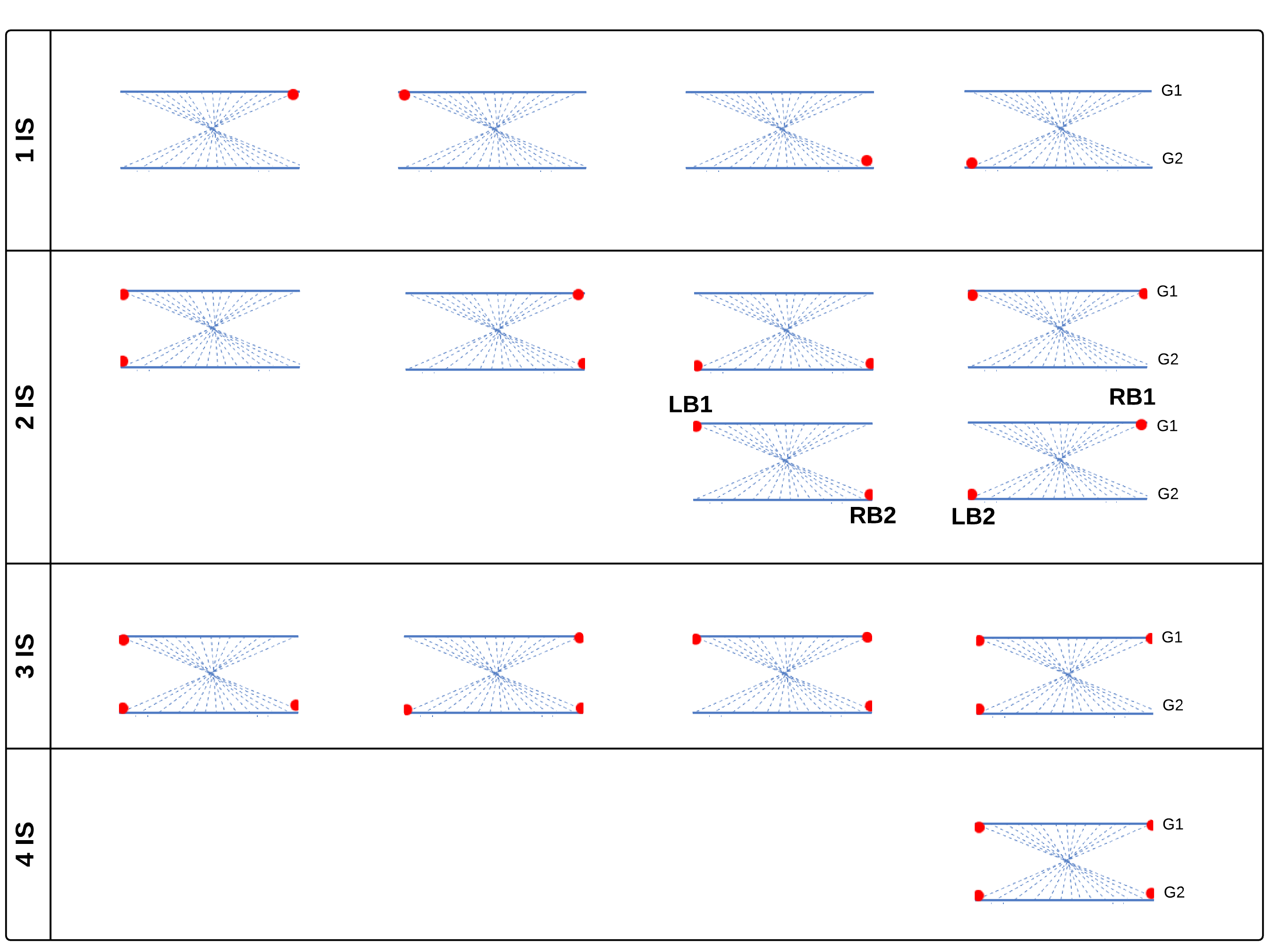}
\caption{Different cases of IS inversions}
\label{fig:cases}
\end{figure}

\item \textbf{Step 5}: Finally, compute the presence probability for each IS families and groups near inversions. (Figure~\ref{fig:cases}).
\end{itemize}

\begin{figure}[!ht]
\centering
\includegraphics[scale=0.5]{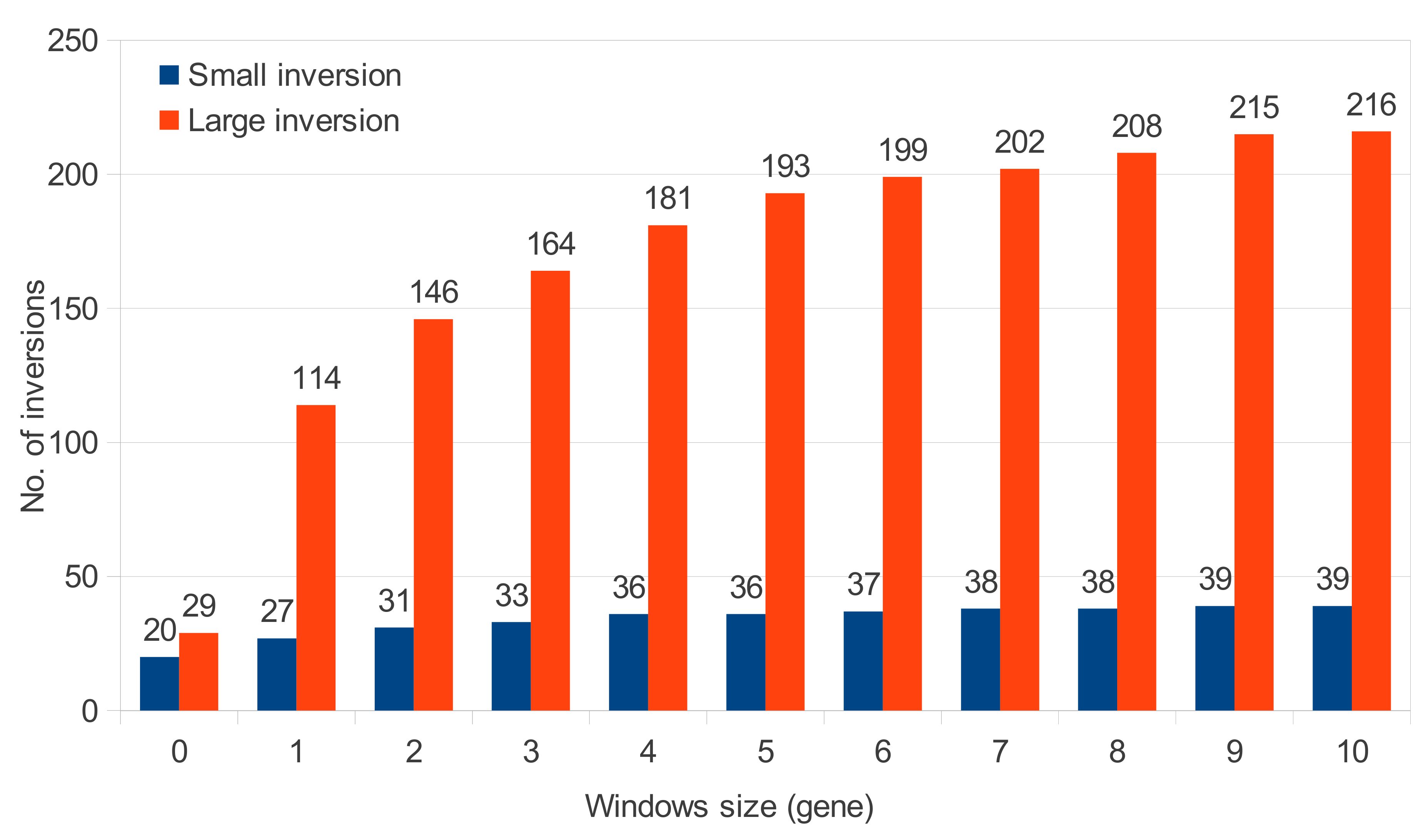}
\caption{Small inversions ($\leqslant$ 4 genes) vs. large inversions ($>4$ genes).}
\label{fig:small_inv}
\end{figure}

As presented in Figure~\ref{fig:small_inv}, there is no major problem in dealing with small 
inversions because the small inversions having small ratio of increment as compared with big 
inversions (\textit{i.e.}, during window size increment of inversion boundaries, the  
small inversions, which have length lower than 4 genes, have small increase ratios compared to large inversions).     

Table~\ref{tab:inv_info} details the roles of IS in largest inversions 
found within two close isolates.

\begin{table}[!ht]
\center
\caption{Summary of large inversion sets within closed genomes}
\scalebox{0.7}{
\begin{tabular}{|c|c|c|c|c|c|c|}
\hline
\textbf{Genome 1 } & \textbf{ Genome 2 } & \textbf{ Inversions no.} & \textbf{ Largest inversion} & \textbf{ IS family} & \textbf{ Boundary} & \textbf{ Window} \\ \hline
19BR  &  213BR  & 9 & 14 & IS110 & LB1-RB2 & w=0\\ \hline
PAO1  &  c7447m  & 2 & 2 & IS3 & (LB1-RB2)/(LB2-RB1) & w=0\\ \hline
LES431  &  LESB58  & 3 & 2 & IS3 & (LB1-RB2)/(LB2-RB1) & w=0\\ \hline
DK2  &  RP73 & 93 & 250 & Tn3 & LB1-RB2 & w=3\\ \hline
UCBPP-PA14  &  B136-33  & 7 & 2 & IS3 & (LB1-RB2)/(LB2-RB1) & w=0\\ \hline
NCGM2.S1  &  MTB-1  & 91 & 255 & IS5 & LB1-RB2 & w=5\\ \hline
\end{tabular}}
\label{tab:inv_info}
\end{table}

\begin{figure}[H]
\center
    \subfloat[Left Boundary\label{fig:tiger}]{%
    \includegraphics[width=0.67\textwidth]{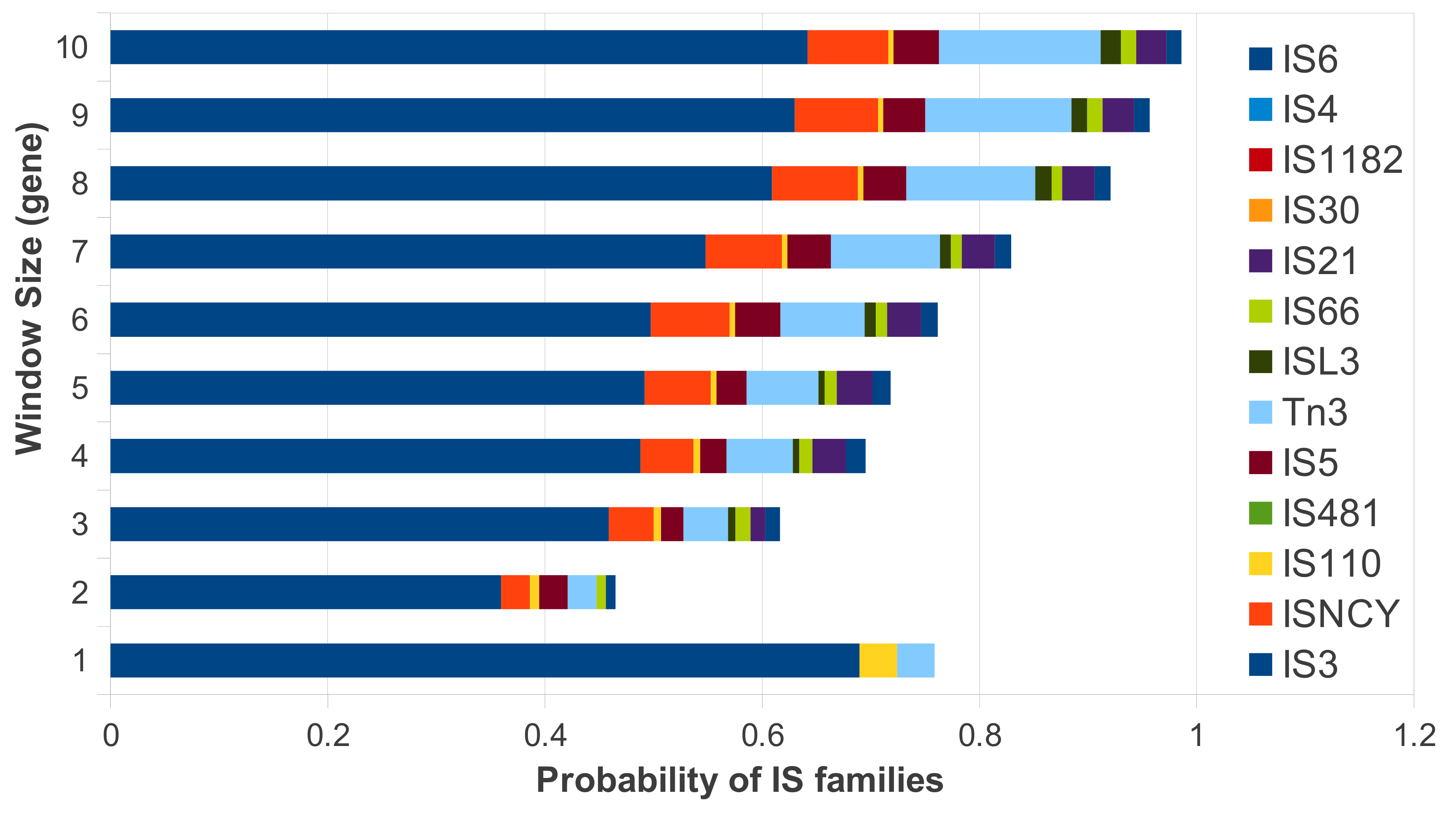}
    }
    \hfill
    \subfloat[Right Boundary.\label{subfig:tb}]{%
      \includegraphics[width=0.67\textwidth]{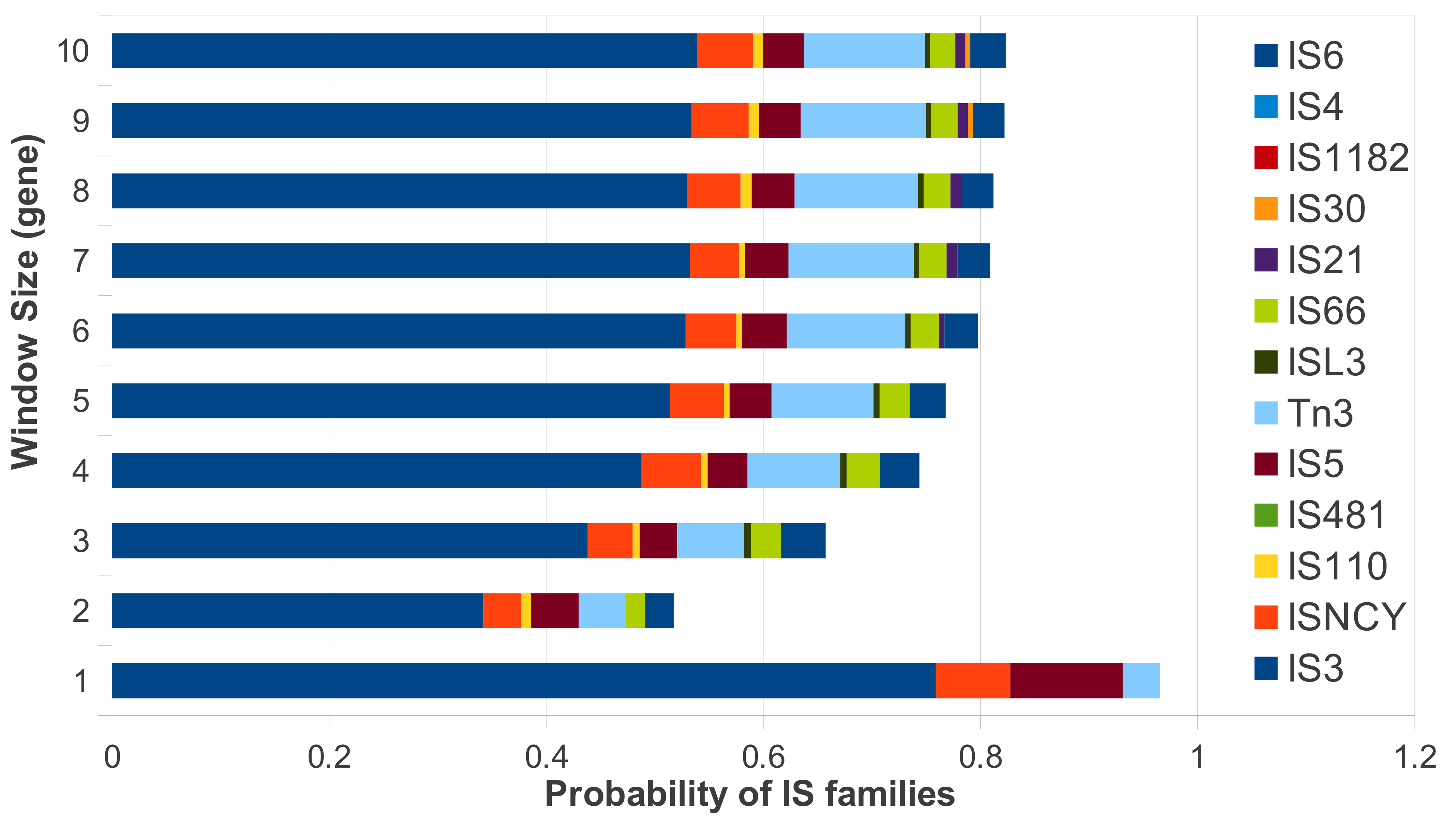}
    }
    
    \caption{IS distribution using different windows size.}
    \label{fig:LB_RB}
\end{figure}

The IS family of type IS3 always have the most probability of appearance with left and right boundaries of inversions.(Figure \ref{fig:LB_RB})

\section{Conclusion}

We designed a pipeline that detects and classify all ORFs that belong to IS. It has been done by merging various tools for ORF prediction, clusterization, and by finally using 
ISFinder database for classification. 

This pipeline has been applied on a set of \textit{Pseudomonas aeruginosa}, showing an obvious improvement in 
ORFs detection that belong to insertion sequences. 
Furthermore, relations between inversions and insertion sequences have been emphasized, leading to the conclusion that
the so-called IS3 family has the largest probability of appearance inside inversion boundaries.

In future works, we intend to investigate more deeply the relation between ISs and other genomic recombination such as deletion and insertion. We will then focus on the implication 
of other types of genes like rRNA (rrnA, rrnB, rrnC, rrnD) in \textit{P. aeruginosa} recombination~\cite{stover2000complete}. By doing so, we will be able to determine genes that are often associated with deletion, inversion, etc. The pipeline will be finally extended to eukaryotic genomes and to other kinds of transposable elements.

\bibliographystyle{unsrt} 
\bibliography{mabase}

\begin{thebibliography}{10}

\bibitem{lin2006spring}
Ying~Chih Lin, Chin~Lung Lu, Ying-Chuan Liu, and Chuan~Yi Tang.
\newblock Spring: a tool for the analysis of genome rearrangement using
  reversals and block-interchanges.
\newblock {\em Nucleic acids research}, 34(suppl 2):W696--W699, 2006.

\bibitem{nref1hawkins2006differential}
Jennifer~S Hawkins, HyeRan Kim, John~D Nason, Rod~A Wing, and Jonathan~F
  Wendel.
\newblock Differential lineage-specific amplification of transposable elements
  is responsible for genome size variation in gossypium.
\newblock {\em Genome research}, 16(10):1252--1261, 2006.

\bibitem{Siguier2006526}
Patricia Siguier, Jonathan Filée, and Michael Chandler.
\newblock Insertion sequences in prokaryotic genomes.
\newblock {\em Current Opinion in Microbiology}, 9(5):526 -- 531, 2006.

\bibitem{citeulike:1766382}
Casey~M. Bergman and Hadi Quesneville.
\newblock {Discovering and detecting transposable elements in genome
  sequences}.
\newblock {\em Briefings in Bioinformatics}, 8(6):382--392, 2007.

\bibitem{kirkpatrick2010and}
Mark Kirkpatrick.
\newblock How and why chromosome inversions evolve.
\newblock {\em PLoS biology}, 8(9):e1000501, 2010.

\bibitem{ranz2007principles}
Jos{\'e}~M Ranz, Damien Maurin, Yuk~S Chan, Marcin Von~Grotthuss, LaDeana~W
  Hillier, John Roote, Michael Ashburner, and Casey~M Bergman.
\newblock Principles of genome evolution in the drosophila melanogaster species
  group.
\newblock {\em PLoS biology}, 5(6):e152, 2007.

\bibitem{hyatt2010prodigal}
Doug Hyatt, Gwo-Liang Chen, Philip~F LoCascio, Miriam~L Land, Frank~W Larimer,
  and Loren~J Hauser.
\newblock Prodigal: prokaryotic gene recognition and translation initiation
  site identification.
\newblock {\em BMC bioinformatics}, 11(1):119, 2010.

\bibitem{van2000graph}
Van Dongen and Stijn Marinus.
\newblock Graph clustering by flow simulation.
\newblock {\em University of Utrecht}, 2000.

\bibitem{siguier2006isfinder}
Patricia Siguier, Jocelyne P{\'e}rochon, L~Lestrade, Jacques Mahillon, and
  Michael Chandler.
\newblock Isfinder: the reference centre for bacterial insertion sequences.
\newblock {\em Nucleic acids research}, 34(suppl 1):D32--D36, 2006.

\bibitem{huda2014}
Huda Al-Nayyef, Christophe Guyeux, and Jacques Bahi.
\newblock A pipeline for insertion sequence detection and study for bacterial
  genome.
\newblock {\em Lecture Notes in informatics (LNI)}, P-235:85--99, 2014.

\bibitem{van2005basys}
Gary~H Van~Domselaar, Paul Stothard, Savita Shrivastava, Joseph~A Cruz, AnChi
  Guo, Xiaoli Dong, Paul Lu, Duane Szafron, Russ Greiner, and David~S Wishart.
\newblock Basys: a web server for automated bacterial genome annotation.
\newblock {\em Nucleic acids research}, 33(suppl 2):W455--W459, 2005.

\bibitem{seemann2014prokka}
T.~Seemann.
\newblock Prokka: rapid prokaryotic genome annotation.
\newblock {\em Bioinformatics}, 30(14):2068--2069, 2014.

\bibitem{robinson2012oasis}
David~G Robinson, Ming-Chun Lee, and Christopher~J Marx.
\newblock Oasis: an automated program for global investigation of bacterial and
  archaeal insertion sequences.
\newblock {\em Nucleic acids research}, 40(22):e174--e174, 2012.

\bibitem{enright2002efficient}
Anton~J Enright, Stijn Van~Dongen, and Christos~A Ouzounis.
\newblock An efficient algorithm for large-scale detection of protein families.
\newblock {\em Nucleic acids research}, 30(7):1575--1584, 2002.

\bibitem{garcia2006mechanism}
Miguel Garcia-Diaz and Thomas~A Kunkel.
\newblock Mechanism of a genetic glissando: structural biology of indel
  mutations.
\newblock {\em Trends in biochemical sciences}, 31(4):206--214, 2006.

\bibitem{hurles2004gene}
Matthew Hurles.
\newblock Gene duplication: the genomic trade in spare parts.
\newblock {\em PLoS biology}, 2(7):e206, 2004.

\bibitem{proost2012adhore}
Sebastian Proost, Jan Fostier, Dieter De~Witte, Bart Dhoedt, Piet Demeester,
  Yves Van~de Peer, and Klaas Vandepoele.
\newblock i-adhore 3.0—fast and sensitive detection of genomic homology in
  extremely large data sets.
\newblock {\em Nucleic acids research}, 40(2):e11--e11, 2012.

\bibitem{Stamatakis21012014}
Alexandros Stamatakis.
\newblock Raxml version 8: A tool for phylogenetic analysis and post-analysis
  of large phylogenies.
\newblock {\em Bioinformatics}, 30(9):1312--1313, 2014.

\bibitem{alkindy2014finding}
Bassam Alkindy, Jean-Fran\c{c}ois Couchot, Christophe Guyeux, Arnaud Mouly,
  Michel Salomon, and Jacques~M. Bahi.
\newblock Finding the core-genes of chloroplasts.
\newblock {\em Journal of Bioscience, Biochemistery, and Bioinformatics},
  4(5):357--364, 2014.

\bibitem{stover2000complete}
CK~Stover, XQ~Pham, AL~Erwin, SD~Mizoguchi, P~Warrener, MJ~Hickey, FSL
  Brinkman, WO~Hufnagle, DJ~Kowalik, M~Lagrou, et~al.
\newblock Complete genome sequence of pseudomonas aeruginosa pao1, an
  opportunistic pathogen.
\newblock {\em Nature}, 406(6799):959--964, 2000.

\end{thebibliography}

\end{document}